\documentclass[11pt]{article}
\usepackage{amsmath,amsfonts,amssymb,graphicx,hyperref,hypcap}
\usepackage{epsfig}
\def\lsim{\mathrel{\rlap{\lower4pt\hbox{\hskip1pt$\sim$}}\raise1pt\hbox{$<$}}}                
\def\gsim{\mathrel{\rlap{\lower4pt\hbox{\hskip1pt$\sim$}}\raise1pt\hbox{$>$}}}  

\def\xvec{{\mathbf{x}}}

\newcommand{\braket}[2]{\langle #1| #2\rangle}

\def\order#1{{\cal O}(#1)}

\begin{document}
\begin{center}
{\Large\bf Semi-classical treatment of $k$-essence effect on cosmic temperature}
\end{center}
\vspace{4mm}
\begin{center}
{\large Abhijit Bandyopadhyay\footnote{Email: abhijit@rkmvu.ac.in}, Debashis Gangopadhyay\footnote{Email: debashis@rkmvu.ac.in} and Arka Moulik\footnote{Email: arkamoulik@rkmvu.ac.in}}
\end{center}

\begin{center}
Department of Physics\\Ramakrishna Mission Vivekananda University\\
Belur Math, Howrah 711202, India
\end{center}
\vspace{4mm}

\begin{abstract}
A phenomenological model is described for Cosmic Microwave Background Radiation 
(CMBR) evolution with dark energy an essential ingredient in the form of a 
$k-$essence scalar field. The following features of this evolution can be 
successfully obtained from this model: (a) the {\it observed} variation of 
the rate of change of scale factor $a(t)$, i.e. $\dot a$, with time and 
(b) the {\it observed} value of the epoch when the  universe went from a 
decelerating phase to an accelerated phase. These two features have been 
matched with  graphical transcriptions of SNe Ia data. The model also 
indicates that the evolution is sensitive to the presence of inhomogeneity
and this sensitivity increases as one goes further into the past. Further,   
the value of the inhomogeneity parameter determines the epoch of switch over 
to an accelerated phase. A positive value of inhomogeneity parameter leads 
to switch over at earlier epochs, while a negative value leads to 
switch over at later epochs. If the value of the inhomogeneity parameter 
is a bit negative then the crossover point from deceleration to 
acceleration gives better agreement with the observed value.
\end{abstract}

\section{Introduction}
\label{sec:1}
Observations of luminosity distances of the type Ia Supernovae (SNe Ia) 
\cite{Riess1,WoodVasey,Davis,Kessler,Kowalski,Amanullah,Xu,
Abraham,simon1,Gaztanaga,Riess2,Stern,paddy} indicate that the universe 
is presently undergoing a phase of accelerated expansion. 
Overwhelming support  exists from other independent observations  
like Cosmic Microwave Background anisotropies measured with WMAP satellite 
\cite{komatsu} and Planck satellite \cite{planckoverview}, 
Baryon Acoustic Oscillations(BAO) \cite {ref:Cole,ref:Huetsi,ref:Percival}
and measurement of oscillations present in the matter power spectrum through 
large scale surveys \cite{eisenstein}.

One of the theoretical approaches in explaining the observed
late time acceleration of the universe is the presence of dark energy
which correspond to a negative pressure in the ideal perfect fluid
model of a Friedmann-Lemaitre-Robertson-Walker (FLRW) universe.
Recent measurements in Planck Satellite experiment \cite{planckoverview} 
suggest dark energy contributes 68.3\% of the total content
of present universe. The issue of origin of dark energy can be addressed 
in the framework of $k-$essence scalar field model of dark energy which 
involve actions with non-canonical kinetic terms. 
In a $k-$essence scalar field model, the kinetic energy 
dominates over the potential energy associated with the scalar field.
Literature on dark energy and $k-$essence models
can be found in \cite{sahn,peebles,malquarti,chimento2,abramo}.

To begin with (Sec.\ \ref{sec:2}), we shall consider a model 
\cite{dg1} where the scalar field is homogeneous, i.e. 
$\phi(t,\xvec)\equiv \phi(t)$ and the FLRW metric has zero curvature constant,
i.e. the universe is flat. A Lagrangian for the $k-$essence 
field that we shall use  has \cite{dg1}  
two generalised coordinates $q(t)= \ln a(t)$ ($a(t)$ is the scale factor) 
and a scalar field $\phi(t)$ with a complicated polynomial interaction 
between them. In this Lagrangian, $q$ has a standard kinetic term 
while $\phi$ does not have a kinetic part and occurs purely through the interaction term.
The general form of $k-$essence Lagrangian is assumed to be a function $L = -V(\phi)F(X)$
with $X= \frac{1}{2}g_{\mu\nu}\nabla^\mu\phi\nabla^\nu\phi$
where $\nabla^{\mu}$ is the covariant derivative, $X$ does not depend explicitly
on $\phi$ to start with and $V(\phi)$ is taken to be a constant.
In \cite{scherrer}, $X$ was shown to satisfy a general
scaling relation, {\it viz.} $X (\frac{dF}{dX})^{2}=Ca(t)^{-6}$ with $C$ a
constant. \cite{dg1} incorporates the scaling relation
of \cite{scherrer}. 

In \cite{dg2} it was shown that the Lagrangian in the above model, 
under certain assumptions reduces to that of a harmonic oscillator 
on the half-plane with time dependent frequency. The quantum mechanical 
amplitude for $q$ to evolve from a value $q_a(t_a)$ to $q_b(t_b)$ 
was computed and using the fact that the scale factor $a(t)$ is 
inversely proportional to the cosmic temperature $T_{a}$ at a given 
epoch $t_{a}$, the quantum amplitude is transformed into an amplitude for evolving from 
$\ln T_{a}$ to $\ln T_{b}$ (Sec.\ \ref{sec:2}).  

In this work we shall show that the above quantum mechanical amplitude is a plausible  phenomenological model of CMBR evolution.

Again, the latest results obtained from the Planck probe 
\cite{planckoverview} have firmly established that inhomogeneity 
effects in CMBR   do  not have their origins in non-gaussianities. 
Hence alternative theoretical approaches to inhomogeneities are desirable.
In this context we shall show that a phenomenological input can be 
introduced in the above model to take into account inhomogeneities. 
This is done by making the scalar field inhomogeneous (Sec.\ \ref{sec:2}).

This phenomenological model has been developed along the following lines 
keeping the observational scenario in mind. First a combined analysis 
of SNe Ia data and Observational Hubble data is done to obtain graphical 
transcriptions of (a) the behaviour of the scale factor $a(t)$ with time
(b) the behaviour of the rate of change of scale factor $\dot a(t)$ 
with time and (c) the second derivative of $a(t)$ with respect to time 
{\it viz.} $\ddot a(t)$.

The observational values thus obtained are then used as inputs in 
the model described as follows. Values of $a(t)$ obtained at specific 
epochs are plugged into the expression for the quantum amplitude to obtain 
the probability profile of the evolution with time. The obtained profile 
is then like the profile of the expectation value of microscopic quantum 
fluctuations, remembering that the expectation value is proportional 
to the transition probability.

The following features of this evolution can be successfully obtained from this model:
\begin{enumerate}
\item[(a)]
The {\it observed} variation of the rate of change 
of scale factor $a(t)$, i.e. $\dot a$, depicted in middle panel of 
Fig.\ \ref{fig:2}, {\it matches with the probability profile obtained 
theoretically from the model} after plugging in observed values of 
$\dot a$ at corresponding epochs, Fig.\ \ref{fig:4}.
\item[(b)]
The {\it observed} value of the epoch when the  universe went from a 
decelerating phase to an accelerated phase, Fig.\ \ref{fig:2}, 
{\it is nearly the same as obtained from the theoretically obtained profile}, 
Fig.\ \ref{fig:4}.
\item[(c)]
There is a qualitative indication that the probability is sensitive 
to the presence of inhomogeneity. This sensitivity increases as one 
goes further into the past. This is seen in Fig.\ \ref{fig:4} and \ref{fig:6}
where the solid line represents homogeneity while the dotted lines 
denote the presence of inhomogeneity.
\item[(d)]
The value of the inhomogeneity parameter determines the epoch of 
switch over to an accelerated phase. A positive value leads to 
switch over at earlier epochs, while a 
negative value leads to switch over at later epochs.
\item[(e)]
If the value of the inhomogeneity parameter is a bit negative then 
the crossover point from deceleration to acceleration gives better 
agreement with the observed value. The homogeneous case (solid line in the figure) 
seems to be roughly the mean curve with respect to positive and 
negative values for the inhomogeneity parameter. This is seen 
in Fig.\ \ref{fig:6}.
\end{enumerate}

\section{Formalism : $k$-essence Lagrangian for scalar field}
\label{sec:2}
We recall briefly the content of references \cite{dg1,dg2}. 
The Lagrangian $L$ (or the pressure $p$) is taken as
\begin{eqnarray}
L &=& -V(\phi) F(X)
\label{eq:1}
\end{eqnarray}
The energy density is
\begin{eqnarray}
\rho &=& V(\phi)[ F(X) -2 X F_{X}]
\label{eq:2}
\end{eqnarray}
with $F_{X}\equiv \frac{dF}{dX}$ and in the present work $V(\phi)=V$ 
is a constant ($> 0$).

For a flat FLRW metric the equation for the $k-$essence field is
\begin{eqnarray}
(F_{X}+2XF_{XX})\ddot\phi+3HF_{X}\dot\phi+(2XF_{X}-F)\frac{V_\phi}{V} &=& 0
\label{eq:3}
\end{eqnarray}
$H = \dot{a}(t)/a(t)$ is the Hubble parameter. Isotropy and  homogeneity 
imply $\phi(x,t)\equiv\phi(t)$, and so $X=\frac{1}{2}\dot\phi^{2}$. 
For $V(\phi)= \mbox{constant}$, one has the scaling law solution \cite{scherrer} 
\begin{eqnarray}
X F_{X}^{2} &=& Ca^{-6}
\label{eq:4}
\end{eqnarray}

Using Eq.\ (\ref{eq:4}), the zero-zero component of Einstein's field equations 
and homogeneity and isotropy an expression 
for the Lagrangian is obtained as \cite{dg1}
\begin{eqnarray}
L=-c_{1}\dot q^{2} - c_{2} V\dot \phi e^{-3q}
\label{eq:5}
\end{eqnarray}
where $a(t)=e^{q(t)}$, $c_{1}= 3(8\pi G)^{-1}$, $c_{2}=2 \sqrt C$,
(we shall always take the positive square root of $C$) 
and the scalar potential $V$ is a constant.

Smaller values of $q$ mean that we are going back to 
smaller values of $a$ i.e. to earlier epochs. Expanding the exponential 
and keeping terms up to $\order{q^{2}}$ one has \cite{dg2}
\begin{eqnarray}
L &=& -\frac{M}{2}\Big{[}\dot q ^{2} + 12\pi G g(t) q^{2}\Big{]} -\frac{1}{2} g(t)
\label{eq:6}
\end{eqnarray} 
where $M=\frac{3}{4\pi G} = \frac{3m_{\mathrm Pl}^{2}}{4\pi}$,
$g(t)= 2{\sqrt C}V \dot\phi$, $m_{\mathrm Pl}$ is the Planck energy
and we use $\hbar = c = 1$ ($c$ is speed of light, $\hbar$ is Planck's constant).
The last term, $\frac{1}{2}g(t)$, is a total derivative.
Dropping this term and the minus sign in front we finally write the Lagrangian as
\begin{eqnarray}
L &=& 
\left(\frac{3}{8\pi G}\right) 
\left[\dot{q}^2 + \left\{24\pi G \sqrt{C}V\dot{\phi} \right\} q^2\right]
\label{eq:7}
\end{eqnarray}
A possible solution for $q$ is obtained when 
$-24\pi G \sqrt{C}V\dot{\phi}$
is a positive number. Writing $\frac{3}{4\pi G} = 
\frac{3m_{\rm pl}^2}{4\pi} \equiv M$
and $-24\pi G\sqrt{C}V\dot{\phi} \equiv \Omega^2(t)$ ($\Omega$ is real),
we rewrite the Lagrangian as
\begin{eqnarray}
L &=& \frac{M}{2}\left[\dot{q}^2 - \Omega^2(t) q^2 \right]
\label{eq:lagosc}
\end{eqnarray}

The harmonic oscillator with time-dependent frequency (Eq.\ \ref{eq:lagosc})   
can now be used as our cosmological model for estimating 
quantum fluctuations of the temperature using path integral technique.
Write the dynamical variable as 
$q(t) = q_{\rm cl}(t) + y(t)$ where $y(t)$ 
($0<y(t)<\infty$) is the fluctuation
over the classical value $q_{\rm cl}(t)$.
This corresponds to a time dependent oscillator in the
half plane \cite{khandlawa,ezawa,simon2,clark}.
Then the quantum mechanical amplitude for $q$ to
evolve from a value $q_a(t_a)$ to $q_b(t_b)$ is given
by \cite{feyn}
\begin{eqnarray}
\braket{q_a,t_a}{q_b,t_b}
&=&
F(t_b,t_a) \exp\left(\frac{i}{\hbar}S_{\rm cl}\right)
\label{eq:9}
\end{eqnarray}
where $S_{cl} = \int_{t_a}^{t_b} L_{\rm cl} dt =
\int_{t_a}^{t_b} dt \frac{M}{2}\left[\dot{q}^2_{\rm cl}
- \frac{1}{2}\Omega^2(t)q_{\rm cl}^2\right]$ and
$F(t_a,t_b)$ is calculated following \cite{khandlawa}. The fluctuations
$y(t)$ satisfy the differential equation
\begin{eqnarray}
\ddot{y} + \Omega^2(t) y &=& 0\,,
\label{eq:y}
\end{eqnarray}
which will have quasi-periodic solutions for real $\Omega$ .
Consider two independent quasi-periodic solutions of  Eq.\ (\ref{eq:y})
\begin{eqnarray}
y_1(t) = \psi(t) \sin\xi(t,t_a)
\quad ; \quad
y_2(t) = \psi(t) \sin\xi(t_b,t)
\label{eq:y1y2}
\end{eqnarray}
with the boundary conditions
\begin{eqnarray}
y_1(t_a) = 0 \quad ; \quad y_2(t_b) = 0
\label{eq:12}
\end{eqnarray}
and where $\psi(t)$ satisfies the Ermakov-Pinney equation \cite{ermakov}
\begin{eqnarray}
\ddot{\psi} + \Omega^2(t) \psi - \psi^{-3} &=& 0
\label{eq:ep}
\end{eqnarray}
with $\xi(t,s)$ defined as
\begin{eqnarray}
\xi(t,s) & \equiv & \nu(t) - \nu(s) 
= \int_{s}^{t} \psi^{-2}(t^\prime)\ dt^\prime 
\label{eq:14}
\end{eqnarray}
$\psi(t)$ and $\nu(t)$ respectively represent the amplitude and phase of
the time dependent oscillator. The fluctuation factor $F(t_b,t_a)$ is then
given by
\begin{eqnarray}
F(t_b,t_a) 
&=&
\left[\frac
{M \sqrt{(\dot{\nu}_a\dot{\nu}_b)} }
{2\pi i \hbar \sin\xi(t_b,t_a)} \right]^{1/2}
\label{eq:fab}
\end{eqnarray}
So the amplitude is (for 
relevant boundary conditions $q(t_a)=q_a , q(t_b)=q_b$),
\begin{eqnarray}
\braket{q_b,t_b}{q_a,t_a}
&=&
\left[\frac
{M \sqrt{(\dot{\nu}_a\dot{\nu}_b)} }
{2\pi i \hbar \sin\xi(t_b,t_a)} \right]^{1/2}
\left[
\exp\left(\frac{iS^{+}_{\rm cl}}{\hbar}\right) 
- \exp\left(\frac{iS^{-}_{\rm cl}}{\hbar}\right) 
\right]
\label{eq:amp}
\end{eqnarray}
where
\begin{eqnarray}
S_{\rm cl}^{\pm}
&=&
\left(
\frac{\dot{\psi}_b q_b^2}{\psi_b}
- \frac{\dot{\psi}_a q_a^2}{\psi_a}
\right)
+ \frac{1}{\sin\xi(t_b,t_a)}
\left[
(\dot{\nu}_b q_b^2 + \dot{\nu}_a q_a^2)\cos\xi(t_b,t_a) \mp 2\sqrt{\dot{\nu}_b\dot{\nu}_a}q_bq_a
\right]
\label{eq:sclpm}
\end{eqnarray}
We assume $\dot{\nu} \ll 1$, i.e. time rate of change of phase is small. 
Also, in a homogeneous universe, the temperature of the background radiation
is inversely proportional to the scale factor i.e. $T(t) \sim \frac{1}{a(t)}$. Then
to lowest order in $\dot{\nu}$, one has the probability for the logarithm
of scale factor or logarithm of inverse temperature evolution as
\begin{eqnarray}
P(t_b,t_a) \equiv P(b,a)
&=&
\left|\braket{q_a,t_a}{q_b,t_b} \right|^2 \nonumber \\
& \equiv &
\left|\left\langle\ln \frac{1}{T_b},t_b \Big{|}
\ln \frac{1}{T_a},t_a \right\rangle
\right|^2 \nonumber \\
&=&
\left(\frac{3m_{\rm pl}^2}{\pi^2\hbar^2 c}\right)
\left(\frac{q_a^2 q_b^2 (\dot{\nu}_a \dot{\nu}_b)^{3/2}}
{\sin^3\xi(t_b,t_a)}\right)  \nonumber \\
&=&
\frac{3m_{\rm pl}^2}{\pi^2} (\ln T_a)^2(\ln T_b)^2
\frac{(\dot{\nu}_a \dot{\nu}_b)^{3/2}}{\sin^3\xi(t_b,t_a)}
\label{eq:18}
\end{eqnarray}
where $\hbar$ and $c$ have been put equal to unity. Expanding the 
function ${(\sin\xi)}^{-3}$ in a Taylor series about $\xi=0$ we  write
\begin{eqnarray}
P(b,a) 
&=&
\frac{3m_{\rm pl}^2}{\pi^2} (\ln T_a)^2(\ln T_b)^2
\frac{(\dot{\nu}_a \dot{\nu}_b)^{3/2}}{\xi^3(t_b,t_a)}
\left[1 + \frac{1}{2}\xi^2(t_b,t_a) 
+ \frac{17}{120}\xi^4(t_b,t_a) + \cdots \right] \nonumber\\
&=&\frac{3m_{\rm pl}^2}{\pi^2} (\ln T_a)^2(\ln T_b)^2
\Bigg{[} p_0(b,a) + p_1(b,a) + p_2(b,a) + \cdots \Bigg{]}
\label{eq:pab}
\end{eqnarray}
where
\begin{eqnarray}
p_0(b,a) = \frac{(\dot{\nu}_a \dot{\nu}_b)^{3/2}}{\xi^3(t_b,t_a)}\,,\quad
p_1(b,a) =\frac{1}{2}
\frac{(\dot{\nu}_a \dot{\nu}_b)^{3/2}}{\xi(t_b,t_a)}\,,\quad
p_2(b,a) = \frac{17}{120}(\dot{\nu}_a \dot{\nu}_b)^{3/2}\xi(t_b,t_a)\,, \cdots
\label{eq:20}
\end{eqnarray}
Choose 
\begin{eqnarray}
\psi(t) = e^{\gamma t}
\quad \mbox{where $\gamma$ is a constant
and $0< \gamma <1$ }
\label{eq:gammahom}
\end{eqnarray}
For this choice of $\psi$,
$\xi(t,s) = \nu(t) - \nu(s) = \int_s^t e^{-2\gamma t} dt = -\frac{1}
{2\gamma}\left[e^{-2\gamma t} -e^{-2\gamma s} \right]$ and therefore,
$\nu(t) = - (1/2\gamma)e^{-2\gamma t}$ ; $\dot{\nu}(t) = e^{-2\gamma t}$.

Using the above expressions for the choice
$\psi(t) = e^{\gamma t}$, and expanding different 
functions as a polynomial of $\gamma$ we obtain
\begin{eqnarray}
(\dot{\nu}_a \dot{\nu}_b)^{3/2}
&=&
e^{-3\gamma (t_b+t_a)} = 
\left[ 1 - 3\gamma (t_b+t_a)  + \frac{9}{2}\gamma^2 (t_b+t_a)^2
+ \cdots \right]
\label{eq:nuadotnubdot} \\
\xi(t_b,t_a) &=& \nu(t_b) - \nu(t_a) = \int_{t_a}^{t_b} e^{-2\gamma t} dt 
\nonumber\\
&=&
(t_b-t_a) \left[1 - \gamma (t_b+t_a) + \frac{2}{3}\gamma^2(t_b^2 + t_bt_a + t_a^2) - \cdots \right] \label{eq:xitbta}
\end{eqnarray}
\begin{eqnarray}
\frac{1}{\xi^3}
&=&
\frac{1}{(t_b-t_a)^3}\Big{[}1 + 3 (t_b+t_a)\gamma +2\left[2t_a^2+2t_b^2 +5 t_bt_a\right]\gamma^2 + \cdots \Big{]}
\label{eq:xi-3ex}\\
\frac{1}{\xi}
&=&
\frac{1}{(t_b-t_a)} 
 \left[1 +(t_b+t_a)\gamma +
\frac{1}{3}\left[t_a^2 + t_b^2 + 4t_at_b \right]\gamma^2 
+\cdots\right]
\label{eq:xi-1ex}
\end{eqnarray}
Using Eqs.\ (\ref{eq:nuadotnubdot}),(\ref{eq:xitbta}),
(\ref{eq:xi-3ex}) and (\ref{eq:xi-1ex})
we calculate the terms $p_0(b,a)$, $p_1(b,a)$, $p_2(b,a) \cdots$ 
appearing in Eq.\ (\ref{eq:pab}) as
\begin{eqnarray}
p_0(b,a) 
&=& 
\frac{(\dot{\nu}_a \dot{\nu}_b)^{3/2}}{\xi^3(t_b,t_a)} 
=
\frac{1}{(t_b-t_a)^3} 
\Bigg{[} 1 - \frac{1}{2} (t_b-t_a)^2\gamma^2 + \cdots 
\Bigg{]}
\label{eq:p0gamma}\\
p_1(b,a) 
&=&
 \frac{1}{2}
\frac{(\dot{\nu}_a \dot{\nu}_b)^{3/2}}{\xi(t_b,t_a)} \nonumber\\
&=&
\frac{1}{2(t_b-t_a)} \Bigg{[} 
1 - 2 \left(t_b+t_a\right)\gamma + \frac{1}{6}
\left(11t_b^2 + 11t_a^2 
+ 26t_at_b\right)\gamma^2
+ \cdots
\Bigg{]} \label{eq:p1gamma}\\
p_2(b,a) 
&=& 
\frac{17}{120}(\dot{\nu}_a \dot{\nu}_b)^{3/2}\xi(t_b,t_a) 
\nonumber \\
&=&
\frac{17}{120}(t_b-t_a) \Bigg{[}
1 - 4 (t_b+t_a)\gamma
+\frac{1}{6}\left(49 t_b^2+ 49 t_a^2 +
94 t_  bt_a\right)\gamma^2 + \cdots \Bigg{]}
\label{eq:p2gamma}
\end{eqnarray}

%
%
The inhomogeneous situation is obtained when relevant quantities have spatial dependence,
i.e. dependence on $\xvec \equiv (r,\theta,\varphi)$, where $(r,\theta,\varphi)$
being the comoving coordinates appearing in the FLRW metric. Then
\begin{eqnarray}
X 
&=& 
\frac{1}{2}g^{\mu\nu}\partial_\mu\phi\partial_\nu\phi \nonumber\\
&=&
\frac{1}{2}\left[ (\partial_t\phi)^2
-\frac{1}{a^2} (\partial_r\phi)^2
- \frac{1}{a^2r^2} (\partial_\theta\phi)^2
- \frac{1}{a^2r^2\sin^2\theta} (\partial_\varphi\phi)^2 \right] 
\end{eqnarray}
Now, for $k$-essence fields the kinetic energy term dominates over the potential 
energy i.e.  $|\partial_{t}\phi|^{2}$  dominates over the square of the $r$,
$\theta$ and $\varphi$  derivatives of the field $\phi$, so that, 
$X \approx \frac{1}{2}\dot{\phi}^2(t,\xvec)$.
We write $\phi(t,\xvec) = \phi(t)\cdot \phi_1(\xvec) = \phi(t)\cdot 
\left(1 + g(\xvec)\right)\sim \phi(t)(1 + f) $,   
where  we have assumed an expansion  
$g(x)= \Sigma f_{n}x^{n}$, $n=0,\cdots,\infty$  with $f_{0}\equiv f$.
The thing to remember is that $f$ is always nonzero and small i.e. 
$0 < | f | < 1$ in presence of small inhomogeneity and the situation of $f$ being zero means homogeneous universe. 
Hence in this work the inhomogeneity  is introduced through a phenomenological 
parameter whose non zero value signifies the presence of inhomogeneity. Here 
the inhomogeneity is introduced through a function which  need not be well behaved 
{\it everywhere} so that it need not be constant. However, 
the first term in the series expansion of this (analytic) function is non-zero, 
more specifically is small i.e. lies between zero and unity. 
This particular  approach will ensure a  phenomenological computation of  
effects of inhomogeneity. In this  work, we are working  out the  
zeroth order (in  spatial dependence) correction. But the  
formalism is sufficiently general to calculate up to higher orders. 
We then have $X = \frac{1}{2}\dot{\phi}^2(1+f)^2$
and for constant $V(\phi)$ the form of the Eq.\ (\ref{eq:3}) still remains the same
and the validity of the scaling relation $XF_X^2 = Ca^{-6}$ is again ensured.
Moreover, in the derivation of Eq.\ (\ref{eq:4}), nowhere was it assumed that 
$\phi$ is homogeneous. The crucial assumption was that $V(\phi)$ is a constant.
So, the presence of inhomogeneity does not change the scaling relation.

Imposing these constraints the Lagrangian Eq.\ (\ref{eq:7}) reduces to
\begin{eqnarray}
L &=& -\sqrt{2C}a^{-3}V (\partial_t \phi (t,\xvec)) 
- \left(\frac{3}{8\pi G}\right)H^2
\label{eq:30}
\end{eqnarray}
where we have dropped a total derivative term as before. 
Writing $\phi(t,\xvec)\sim\phi(t)(1+f)$
and proceeding as in the previous section taking 
$V(\phi)=\mbox{constant}=V$ we have
\begin{eqnarray}
L
&=&
\frac{M}{2}\left[ \dot{q}^2 
+\left\{24\pi G \sqrt{C}V\dot{\phi} \right\}
\Big{(}1 + f \Big{)} q^2\right] \nonumber\\
&=& \frac{M}{2}\left[ \dot{q}^2 - \Omega^2(t) 
\Big{(}1 + f \Big{)} q^2\right]\nonumber\\
&=&\frac{M}{2}\left[ \dot{q}^2 - \Omega^2_f(t) q^2\right]
\end{eqnarray}
where $\Omega_f(t) \equiv (1 + f)^{1/2} \Omega(t)$.

It is but natural that the scale factor cannot take the same value when 
inhomogeneity is present. Technically this means that {\it now it has to be some 
different function of the time $t$}. To distinguish these two cases, i.e. homogeneous and 
inhomogeneous, we write the scale factor in the presence of inhomogeneity as 
$a_{f} (t)$ instead of $a(t)$ which denotes the homogeneous scenario. 
In the same spirit, the functions  $\nu_{f}, \psi_{f}, 
\xi_{f}$ are also different functions of time from their homogeneous counterparts , 
{\it viz.}, $\nu, \psi$ and $\xi$.

To estimate quantum fluctuations of the temperature in the presence 
of inhomogeneity ($f$), we first assert that the dynamical variable
$q(t)=\ln a(t)$ will be modified in presence of inhomogeneity
which we denote by the notation $q_f(t)=\ln a_f(t)$.
Then we write this as
$q_f(t) = q_{f, \rm cl}(t) + y_f(t)$ where $y_f(t)$ 
($0<y_f(t)<\infty$) is the fluctuation
over the classical value $q_{f, \rm cl}(t)$.
We take $q_{f,\rm cl} = q_{\rm cl}$, 
so that $q_f(t) = q_{\rm cl}(t) + y_f(t)$ and $y_f(t)$ 
gives the fluctuation (in presence of inhomogeneity) 
over the same classical value $q_{\rm cl}$. 
Then the quantum mechanical amplitude 
for $q_f$ to evolve from a value $q_{fa}(t_a)$ to  
$q_{fb}(t_b)$ is given by
\begin{eqnarray}
\braket{q_{fa},t_a}{q_{fb},t_b}
&=&
F_f(t_b,t_a) \exp\left(\frac{i}{\hbar}S_{f,\rm cl}\right)
\label{eq:32}
\end{eqnarray}
where $S_{f,\rm cl} = \int L_{f,\rm cl} dt = \int_{t_a}^{t_b} 
\frac{M}{2}\Big{[}\dot{q}_{\rm cl}^2 - \Omega_f^2(t)q_{\rm cl}^2 \Big{]}$ 
which is structurally same as $S_{\rm cl}$ but with $\Omega$ replaced by
$\Omega_f$; and the fluctuation factor $F_f(t_a,t_b)$ can again be calculated 
following \cite{khandlawa}. The fluctuations $y_f(t)$ now satisfies the differential
equation
\begin{eqnarray}
\ddot{y}_f + \Omega^2_f(t) y_f &=& 0\,.
\label{eq:yf}
\end{eqnarray}
As before, two independent quasi-periodic solutions of Eq.\ (\ref{eq:yf})
(for real $\Omega_f$) can be considered as
\begin{eqnarray}
y_{f1}(t) = \Psi_f(t)\sin\xi_f(t,t_a) \quad , \quad
y_{f2}(t) = \Psi_f(t)\sin\xi_f(t_b,t)
\end{eqnarray}
with boundary conditions 
\begin{eqnarray}
y_{f1}(t_a) = 0 \quad ; \quad y_{f2}(t_b) = 0
\label{eq:35}
\end{eqnarray}
where, $\Psi_f(t)$ satisfies the Ermakov-Pinney equation
\begin{eqnarray}
\ddot{\Psi}_f + \Omega^2_f(t)\Psi_f - \Psi_f^{-3} &=& 0
\label{eq:epf}
\end{eqnarray}
with $\xi_f(t,s)$ defined as 
\begin{eqnarray}
\xi_f(t,s) \equiv \nu_f(t) - \nu_f(s) 
= \int_s^t \Psi_f^{-2}(t^\prime)dt^\prime
\end{eqnarray}
Here $\Psi_f(t)$ and $\nu_f(t)$ respectively represent
the amplitude and phase of the time dependent oscillator. 
The fluctuation factor $F_f(t_b,t_a)$ in presence of 
inhomogeneity is then given by
\begin{eqnarray}
F_f(t_b,t_a) 
&=&
\left[\frac
{M \sqrt{\dot{(\nu_f)}_a\dot{(\nu_f)}_b} }
{2\pi i \hbar \sin\xi_f(t_b,t_a)} \right]^{1/2}
\label{eq:ffab}
\end{eqnarray}
So the amplitude is (for relevant boundary conditions $q_f(t_a)=q_{fa} , q_f(t_b)=q_{fb}$),
\begin{eqnarray}
\braket{q_{fb},t_{fb}}{q_a,t_a}
&=&
\left[\frac
{M \sqrt{\dot{(\nu_f)}_a\dot{(\nu_f)}_b} }
{2\pi i \hbar \sin\xi_f(t_b,t_a)} \right]^{1/2}
\Big{[}
\exp\left(\frac{iS^{+}_{f,\rm cl}}{\hbar}\right) 
- \exp\left(\frac{iSf^{-}_{f,\rm cl}}{\hbar}\right) 
\Big{]}
\label{eq:ampf}
\end{eqnarray}
where
\begin{eqnarray}
S_{f,{\rm cl}}^{\pm}
&=&
\left(
\frac{\dot{\Psi}_{fb} q_{fb}^2}{\Psi_{fb}}
- \frac{\dot{\Psi}_{fa} q_{fa}^2}{\Psi_{fa}}
\right) \nonumber\\
&&
+ \frac{1}{\sin\xi_f(t_b,t_a)}
\Big{[}
(\dot{\nu}_{fb} q_{fb}^2 + \dot{\nu}_{fa} q_{fa}^2)\cos\xi_f(t_b,t_a) \mp 2\sqrt{\dot{\nu}_{fb}\dot{\nu}_{fa}}q_{fb}q_{fa}
\Big{]}
\label{eq:sclpmf}
\end{eqnarray}
For small $\dot{\nu}_f \ll 1$, 
to lowest order in $\dot{\nu}$, the probability of transition for logarithm
of scale factor (inverse temperature) in presence of
inhomogeneity would be
\begin{eqnarray}
P_f(b,a)
&=&
\left|\braket{q_{fa},t_a}{q_{fb},t_b} \right|^2 =
\frac{3m_{\rm pl}^2}{\pi^2} (\ln T_{fa})^2(\ln T_{fb})^2
\frac{(\dot{\nu}_{fa} \dot{\nu}_{fb})^{3/2}}{\sin^3\xi_f(t_b,t_a)}
\label{eq:pbaf}
\end{eqnarray}

We now choose to write
\begin{eqnarray}
\Psi_f(t) &=& (1 + f) \psi(t)
\label{eq:phif}
\end{eqnarray}
so that for $f = 0$, $\Psi_f(t) \to \psi(t)$  and $\Omega_f(t) \to \Omega(t)$
and we get back homogeneous scenario. Also for such a choice the quantities 
$\nu_f$ and $\xi_f$ appearing in Eq.\ (\ref{eq:pbaf}) can be expressed 
explicitly in terms of respective quantities $\nu$ and $\xi$ 
corresponding to the homogeneous scenario in manner described below.

If we choose $e^{\gamma_f(t)}$ to represent a solution of the 
Ermakov-Pinney equation [Eq.\ (\ref{eq:epf})], then in order the
Eq.\ (\ref{eq:phif}) is satisfied, $\gamma_f(t)$ will be related to
the parameter $\gamma$ (recall that: $e^{\gamma t}$ represents a solution of the
Ermakov-Pinney equation (Eq.\ (\ref{eq:ep})) corresponding to 
the homogeneous case) by the following relation
\begin{eqnarray}
\gamma_f(t) &=& \gamma t + \ln (1 + f) 
\end{eqnarray}

Note here that the way we choose to incorporate the inhomogeneity 
into our scheme retains the Ermakov-Pinney structure of the differential 
equations which is the central feature of the formalism described in Sec. \ref{sec:2}. 
This will ensure that new solutions thereby obtained will obey 
all the mathematical properties of Ermakov theory regarding 
the form of the nonlinear equations.

It is quite evident that $\Psi$ reduces to $\psi(t)$ if $f = 0$
and then $\gamma_f(t)$ reduces to $\gamma t$ and we
get back the original situation. Therefore we can unambiguously use 
the previous formalism with $P(b,a)$ replaced by  $P_f(b,a)$,
$\nu(t)$'s replaced by $\nu_f(t)$'s and $\xi(t_b,t_a)$ 
replaced by $\xi_f(t_b,t_a)$ respectively.
Now with $\Psi_f(t) = e^{\gamma_f(t)}$
the new $\nu_f(t)$ and $\xi_f(t_b,t_a)$ are related to 
corresponding functions for homogeneous case by 
the following relations
\begin{eqnarray}
\nu_f(t) &=& \frac{\nu(t)}{[1 + f]^2}
\label{eq:numod} \\
\xi_f(t_b,t_a) 
&=&
\nu_f(t_b) - \nu_f(t_a) = \frac{\nu(t_b) - \nu(t_a)}{[1 + f]^2}
 \label{eq:ximod}
\end{eqnarray}
Using Eqs.\ (\ref{eq:numod}) and (\ref{eq:ximod}) 
in Eq.\ (\ref{eq:pbaf}) we get
\begin{eqnarray}
P_f(b,a)
&=&
\frac{3m_{\rm pl}^2}{\pi^2} (\ln T_a)^2(\ln T_b)^2
\Bigg{[} p_{0f}(b,a) + p_{1f}(b,a) + p_{2f}(b,a) + \cdots \Bigg{]}
\label{eq:pabphi}
\end{eqnarray}
where the quantities $p_{0f}$, $p_{1f}$, $p_{2f}, \cdots$ appearing
in the above equation is expressible in terms of their homogeneous
counterparts as
\begin{eqnarray*}
p_{0f}(b,a) &=& \frac{[\dot{\nu}_f(t_a) \dot{\nu}_f(t_b)]^{3/2}}{\xi^3_f(t_b,t_a)} 
=  
\frac{[\dot{\nu}(t_a) \dot{\nu}(t_b)]^{3/2}/(1+f)^6}
{\xi^3(t_b,t_a)/(1+f)^6} \\
&=& p_0(b,a)\\
p_{1f}(b,a) &=& \frac{1}{2}
\frac{[\dot{\nu}_f(t_a) \dot{\nu}_f(t_b)]^{3/2}}{\xi_f(t_b,t_a)} 
=
 \frac{1}{2}
\frac{[\dot{\nu}(t_a) \dot{\nu}(t_b)]^{3/2}/(1+f)^6}{\xi(t_b,t_a)/(1+f)^2} \\
&=& 
\frac{p_1(b,a)}{(1+f)^4} = p_1(b,a)(1 - 4f + \order{f^2})
\\
p_{2f}(b,a) &=& \frac{17}{120}[\dot{\nu}_f(t_a) \dot{\nu}_f
(t_b)]^{3/2}\xi_f(t_b,t_a) \\
&=&
\frac{p_2(b,a)}{(1+f)^8} = p_2(b,a)(1 - 8f + \order{f^2})\\
&\vdots &\\
p_{nf}(b,a) &=& \frac{p_n(b,a)}{(1+f)^{4n}} = p_n(b,a) (1 - 4nf + \order{f^2})
\end{eqnarray*}
In terms of $\gamma$, $t_a$ and $t_b$ the above expression becomes
\begin{eqnarray}
P_f(b,a)
&=&
\frac{3m_{\rm pl}^2}{\pi^2} (\ln T_a)^2(\ln T_b)^2 \times \nonumber\\
&&
\Bigg{[}
\frac{1}{(t_b-t_a)^3} 
\Big{[} 1 - \frac{1}{2} (t_b-t_a)^2\gamma^2 + \cdots 
\Big{]} \nonumber\\
&&
+ 
\frac{1}{2(t_b-t_a)(1 + f)^4} \Big{[} 
1 - 2 \left(t_b+t_a\right)\gamma + \frac{1}{6}
\left(11t_b^2 + 11t_a^2 
+ 26t_at_b\right)\gamma^2
+ \cdots
\Big{]} \nonumber\\
&&
+ 
\frac{17}{120}\frac{(t_b-t_a)}{(1+f)^8} \Big{[}
1 - 4 (t_b+t_a)\gamma
+\frac{1}{6}\left(49 t_b^2+ 49 t_a^2 +
94 t_  bt_a\right)\gamma^2 + \cdots \Big{]}
\Bigg{]} 
\label{eq:expand}
\end{eqnarray}

\section{Extracting time dependence of scale factor from
observational data}
\label{sec:3}
%
\begin{figure}
\begin{center}
\includegraphics[width=8.5cm, height=7cm]{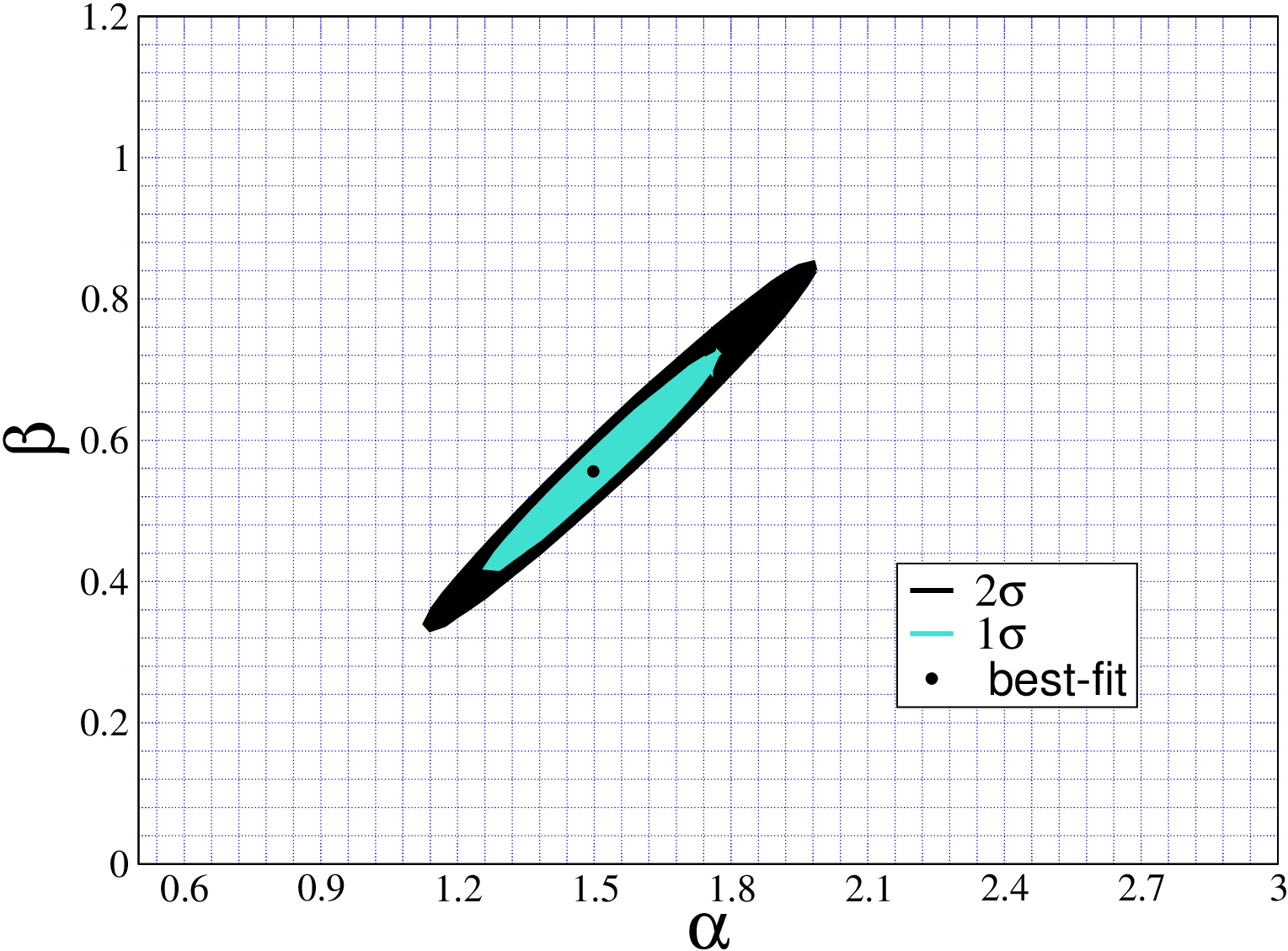}
\caption{Allowed regions in the parameter space $\alpha-\beta$ at 
1$\sigma$ and 2$\sigma$ confidence level from the combined analysis of SNe Ia data and OHD}
\label{fig:1}
\end{center}
\end{figure}
Measurement of luminosity distance of the type Ia Supernovae (SNe Ia)
during nearly last two decades establishes that the universe is 
presently undergoing a phase of accelerated expansion. 
Observation of Baryon Acoustic Oscillations (BAO), 
Cosmic Microwave Background (CMB) radiations, power spectrum
of matter distributions in the universe provide other independent 
evidence in favour of this late-time cosmic acceleration. 
However, the SNe Ia data remain the  key observational 
ingredient in determining time evolution of the scale factor $a(t)$
in the late-time phase of evolution of the universe. 
Besides the SNe IA data, observational data based on measurement 
of differential ages of the galaxies by Gemini Deep Deep Survey 
GDDS \cite{Abraham}, SPICES  and VDSS surveys also provide 
dependence of Hubble parameter with redshift.
We have extracted the  this time evolution 
from the combined analysis of SNe Ia data and observational 
Hubble data.

\subsection{Methodology of analysis of SNe Ia and Observational Hubble data}
Here we briefly describe the methodology we use for the combined
analysis of SNe IA and Observational Hubble Data (OHD).
We use a closed form parametrisation of the 
luminosity distance of supernova, $d_L$, as a function of the redshift as \cite{paddy}
\begin{eqnarray}
d_L(\alpha,\beta,z) &=& \frac{c}{H_0} \left(\frac{z(1 + \alpha z)}{1+ \beta z}\right)
\label{eq:dL1}
\end{eqnarray}
where $c$ is the speed of light and $H_0$ the value of the 
Hubble  parameter at the present epoch defined through the dimensionless
quantity $h$ by $H_0 = 100\ h\ $km s$^{-1}$ Mpc$^{-1}$.
The luminosity distance is related to the distance modulus
$\mu$ as
\begin{eqnarray}
\mu_{\rm th}(\alpha,\beta,z) 
&=& 5\log_{10}\Big{[}D_L(\alpha,\beta,z)\Big{]} + \mu_0 \, \nonumber\\
&=& 5\log_{10}\Big{[}\left(\frac{z(1 + \alpha z)}{1+ \beta z}\right)\Big{]} + \mu_0
\end{eqnarray}
where,
\begin{eqnarray}
D_L(\alpha,\beta,z) \equiv \frac{H_0}{c}d_L(\alpha,\beta,z)
= \left(\frac{z(1 + \alpha z)}{1+ \beta z}\right)
\label{eq:DL}
\end{eqnarray}
is a dimensionless quantity called the Hubble free luminosity distance
and $\mu_0 = 42.38 - 5\log_{10} h$.
From different compilations of SNe Ia observations by different groups - 
HST + SNLS + ESSENCE \cite{Riess1,WoodVasey,Davis},
SALT2 and MLCS data \cite{Kessler}, UNION \cite{Kowalski} and
UNION2 data \cite{Amanullah}.
provide the values of the distance modulus 
for different values of the redshift from the SNe Ia observations. 
The observed values of the distance modulus $\mu_{\rm obs}(z_i)$ 
corresponding to  measured redshifts $z_i$ are given in terms of the 
absolute magnitude $M$ and the apparent magnitudes $m_{\rm obs}(z_i)$ by
\begin{eqnarray}
\mu_{obs}(z_i)=m_{obs}(z_i)-M\,.
\end{eqnarray} 
To obtain the best-fit values of the parameters $\alpha$ and $\beta$ from 
SNe Ia observations we perform a likelihood analysis whose methodology
has been discussed in detail in \cite{Xu}. This
involves minimization of a suitably chosen $\chi^2$ function 
with respect to the 
parameters $\alpha$ and $\beta$. We give below a brief outline
of the methodology of $\chi^2$-analysis adopted here for the analysis of SNe Ia data.
The $\chi^2$ function is defined as the function of the parameters
$\alpha$, $\beta$ and $M^\prime\equiv \mu_0 +M$ (called nuisance parameter)
as
\begin{eqnarray}
\chi^2(\alpha,\beta,M^{\prime}) 
&=& \sum_{i=1}^N\frac{(\mu_{obs}(z_i)-
\mu_{th}(\alpha,\beta,z_i))^2}{\sigma_i^2} \nonumber\\
&=& \sum_{i=1}^N\frac{\Big{[}5\log_{10}D_L(\alpha,\beta,z_i)-m_{obs}
(z_i)+M^{\prime}\Big{]}^2}{\sigma_i^2} \quad
\end{eqnarray}
where $\sigma_i$'s are the uncertainties in observations of distance modulus 
$\mu_{obs}(z_i)$'s, and  $N$ is the total number of data points.
The values of the parameters $\alpha$ and $\beta$ (appearing
in parametrisation of luminosity distance) which fits the
SNe IA data best, are those  which minimizes the $\chi^2$ function after the 
parameter $M^\prime$ is marginalised over. 
Expanding the $\chi^2$ function as
\begin{eqnarray}
\chi^2 &=& P(\alpha,\beta) + 2Q(\alpha,\beta) M^\prime + R{M^\prime}^2
\end{eqnarray}
where
\begin{eqnarray}
\displaystyle P(\alpha,\beta) &=& \sum_{i=1}^N\frac{(5\log_{10}
(D_L(\alpha,\beta,z_i))-m_{obs}(z_i))^2
}{\sigma_i^2} \label{eq:P}\\
\displaystyle Q(\alpha,\beta)  &=& \sum_{i=1}^N\frac{(5\log_{10}
(D_L(\alpha,\beta,z_i)) - m_{obs}(z_i))
}{\sigma_i^2} \label{eq:Q}\\
\displaystyle R &=&\sum_{i=1}^N\frac{1}{\sigma_i^2} \label{eq:R}
\end{eqnarray}
we observe that, the $\chi^2$ have a minimum at $M^\prime = -Q/R$ and its value at
the minimum is $\bar{\chi}^2 (\alpha,\beta) = P - Q^2/R$. To obtain the best-fit value 
of the parameters $\alpha$ and $\beta$ its then enough to minimize the function
$\bar{\chi}^2 (\alpha,\beta)$ with respect to $\alpha$ and $\beta$ 
only since the effect of marginalisation over $M^\prime$
gets taken care of in the above consideration. So the $\chi^2$-function
for analysis of SNe IA data used here is
\begin{eqnarray}
\chi_{\rm SN}^2(\alpha,\beta) & = & P(\alpha,\beta) - \frac{Q^2(\alpha,\beta)}{R}
\label{chifinal}
\end{eqnarray}
where $P(\alpha,\beta)$, $Q(\alpha,\beta)$ and $R$ are given by 
Eqs.\ (\ref{eq:P}),\ (\ref{eq:Q}) and \ (\ref{eq:R}) respectively.

Besides the SNe IA data, compilation of the observational data based on measurement of 
differential ages of the galaxies by Gemini Deep Deep Survey GDDS \cite{Abraham}, SPICES  and 
VDSS surveys provide the values of the Hubble parameter at 15  different redshift values 
\cite{simon1,Gaztanaga,Riess2,Stern}. The $\chi^2$ function for the analysis of this 
observational Hubble data can be defined as
\begin{eqnarray}
\chi^2_{\rm OHD}(\alpha,\beta) &=& \sum_{i=1}^{15} \left [ 
\frac {H(\alpha,\beta;z_i) - H_{\rm obs}(z_i)} {\Sigma_i} \right ]^2 \,\,,
\end{eqnarray}
where $H_{\rm obs}$ is the observed Hubble parameter value at $z_i$ 
with uncertainty $\Sigma_i$.\\

Varying the parameters $\alpha$ and $\beta$ freely we minimize the global $\chi^2$ function which is defined as 
\begin{eqnarray}
\chi^2(\alpha,\beta) &=& \chi^2_{\rm SN}(\alpha,\beta)  
+  \chi^2_{\rm OHD}(\alpha,\beta)\,\,.
\end{eqnarray}
The values of the parameters $\alpha$ and $\beta$ at which minimum of 
$\chi^2$ is obtained are the best-fit values of these parameters for the 
combined analysis of the observational data from SNe Ia and Observational Hubble
Data (OHD). We also 
find the 1$\sigma$ and 2$\sigma$ ranges  of the parameters $\alpha$ and 
$\beta$ from the analysis of the observational data discussed above. In this 
case of two parameter fit, the  1$\sigma$ (68.3\% confidence level) and 
2$\sigma$(95.4\% confidence level) allowed ranges of the parameters 
correspond to $\chi^2 \leq \chi^2_{\rm min} + \Delta\chi^2$, where 
$\Delta\chi^2=2.30(6.17)$ denotes the 1$\sigma$(2$\sigma$) spread in 
$\chi^2$ corresponding to two parameter fit.

In this work we have considered the SNe Ia data from HST+SNLS+ESSENCE 
(192 data points)
\cite{Riess1,WoodVasey,Davis} and Observational Hubble Data from 
\cite{simon1,Gaztanaga,Riess2,Stern} (15 data points). 
The best fit for the combined analysis of the SNe Ia data and OHD 
is obtained for the parameters values
\begin{eqnarray}
\alpha =1.50 \quad , \quad \beta=0.55
\label{eq:ab}
\end{eqnarray}
with a minimum $\chi^2$ of 204.94. 
In Fig.\ \ref{fig:1}
we have shown the regions of the $\alpha-\beta$ parameter space
allowed at $1\sigma$ and $2\sigma$ confidence levels from the analysis.

\subsection{Methodology of obtaining time dependence of scale factor 
from observational data}
\begin{figure}[t]
\centering
\includegraphics[width=\textwidth, height=2.7in]{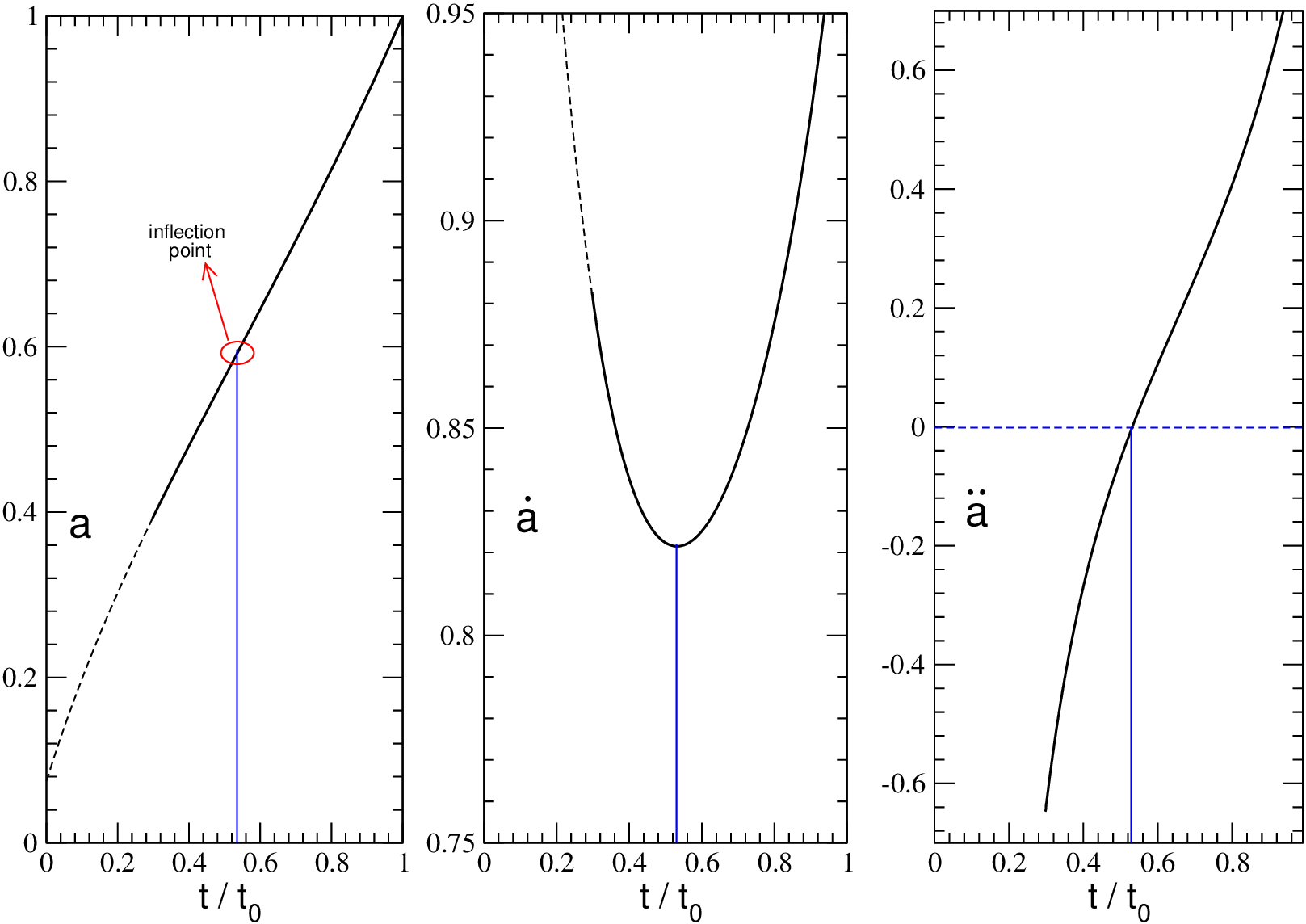}
\caption[]
{Plot of $a$ (left panel), $\dot{a}$ (middle panel)
and $\ddot{a}$ (right panel) against $t$ corresponding to
the best-fit values of parameters $\alpha$ and $\beta$ obtained from
analysis of SN data.}
\label{fig:2}
\end{figure}
Using the values of $\alpha$ and $\beta$ (Eq.\ (\ref{eq:ab}))
as obtained from the analysis we can determine the time dependence of the 
scale factor and the Hubble parameter during the late time evolution of the
universe.

For a flat universe, which is consistent with
the current bounds from PLANCK and WMAP data
on the ratio of the energy density in curvature to the critical density,
$|\Omega_K^{0}| < 0.01$ (95\% confidence level) (PLANCK)
the Hubble parameter $H(z)$ corresponding to a redshift $z$ is directly
related to the luminosity distance through the relation
\begin{eqnarray}
E(z) 
&\equiv & \frac{H(z)}{H_0} 
= \left[\frac{d}{dz}\left( \frac{D_L(z)}{1+z}\right)\right]^{-1}
\label{eq:Ez}
\end{eqnarray}
From the equations $H = \displaystyle\frac{\dot{a}}{a}$ and 
$\displaystyle\frac{a_0}{a} = 1 + z$
we get 
\begin{eqnarray}
dt = -\frac{dz}{(1+z)H} = -\frac{dz}{(1+z)H_0 E(z)}
\end{eqnarray}
which on integration gives
\begin{eqnarray}
\frac{t(z)}{t_0} 
&=& 
1 - \frac{1}{H_0t_0} \int_z^0 \frac{dz^\prime}{(1+z^\prime)E(z^\prime)}
\label{eq:tzdep}
\end{eqnarray}
where $t_0$ is the time corresponding to present epoch. Taking the
best-fit values of $\alpha$ and $\beta$, 
we use Eq.\ (\ref{eq:DL}) to numerically evaluate $D_L(z)$
at different $z$ values. Using this in Eq.\ (\ref{eq:Ez})
we then evaluate 
$E(z)$ as a function of $z$ which can further be used in 
Eq.\ (\ref{eq:tzdep}) to perform the integration numerically
to obtain time $t$ as a function of redshift $z$. From the 
$z - t(z)$ relationship thus obtained and the equation
$a_0/a = 1+z$ we eliminate $z$ to obtain the scale factor
$a$ as a function of $t$.

In Fig.\ (\ref{fig:2})  we have shown the time dependence of the scale factor
as obtained from the analysis of the observational data following technique
described above. The left panel shows plot of $a(t)$ vs $t/t_0$ where the 
value of scale factor at present epoch has been normalised to unity $a(t_0)=1$.
The observed supernova Ia events have redshifts ranging between $0 <z \lsim 1.6 $
which correspond to the range $0.3 \lsim t/t_0 <1$. In the obtained 
$t$-dependence of $a(t)$ there exists a point of inflection
at $t/t_0 \approx 0.53$ ($z \approx 0.68 $).  This feature
transforms into a minimum in the middle panel  where we have plotted $\dot{a}$
against $t$. In the  right panel ($\ddot{a}$ vs $t$) again there is the signature of 
crossover to an accelerated phase of expansion.

\section{Results and discussions}
\begin{figure}[t]
\centering
\includegraphics[width=.7\textwidth, height=2.7in]{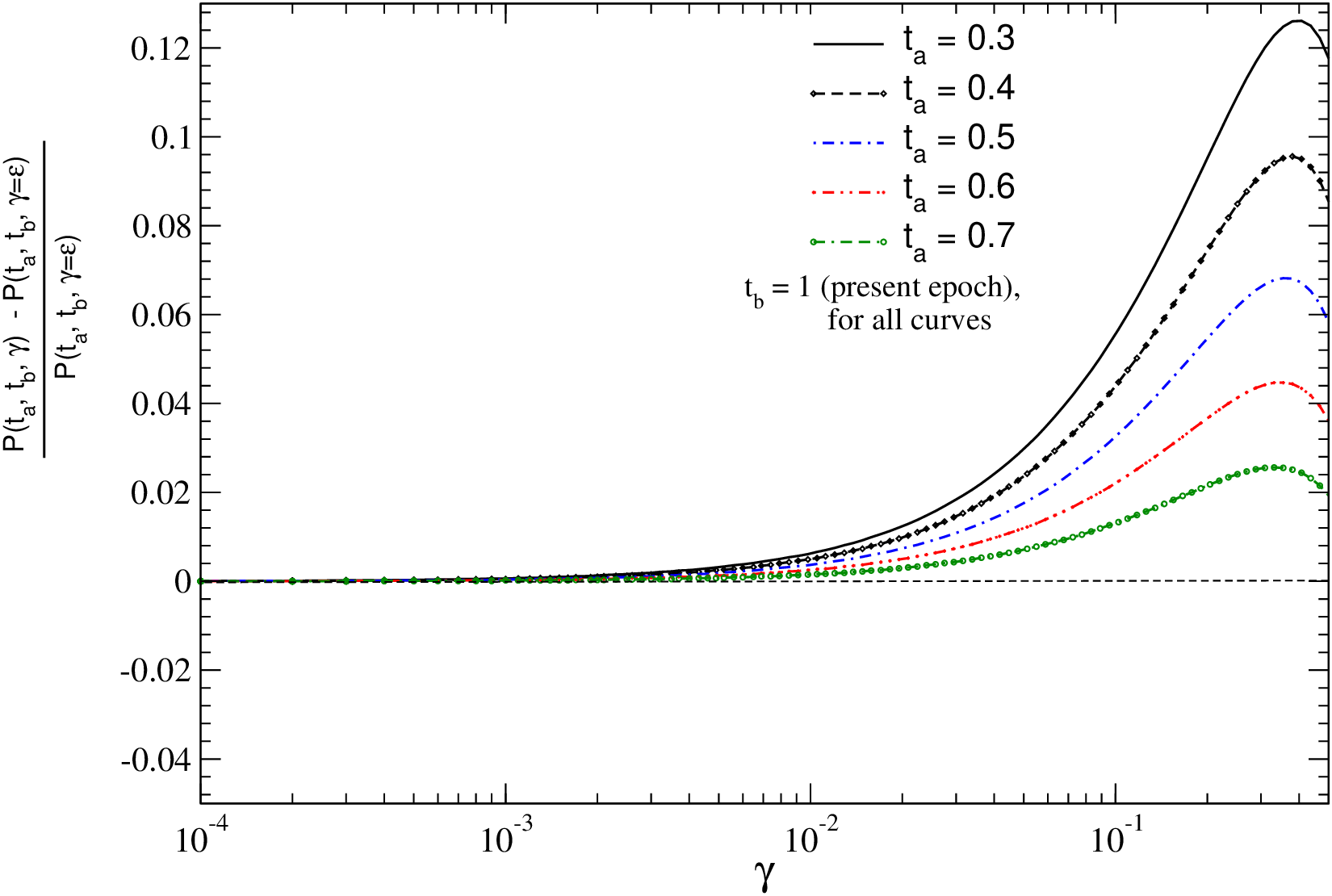}
\caption[]
{Plot of $[P_\gamma(t_a,t_b) - P_{\gamma\to \epsilon}(t_a,t_b)]/
P_{\gamma\to \epsilon}(t_a,t_b)$ vs $\gamma$ for $t_b=1$ (present epoch) 
and for different chosen values of $t_a$. We take $\epsilon=10^{-6}$.}
\label{fig:3}
\end{figure}
%
\begin{figure}[t]
\centering
\includegraphics[width=\textwidth, height=2.7in]{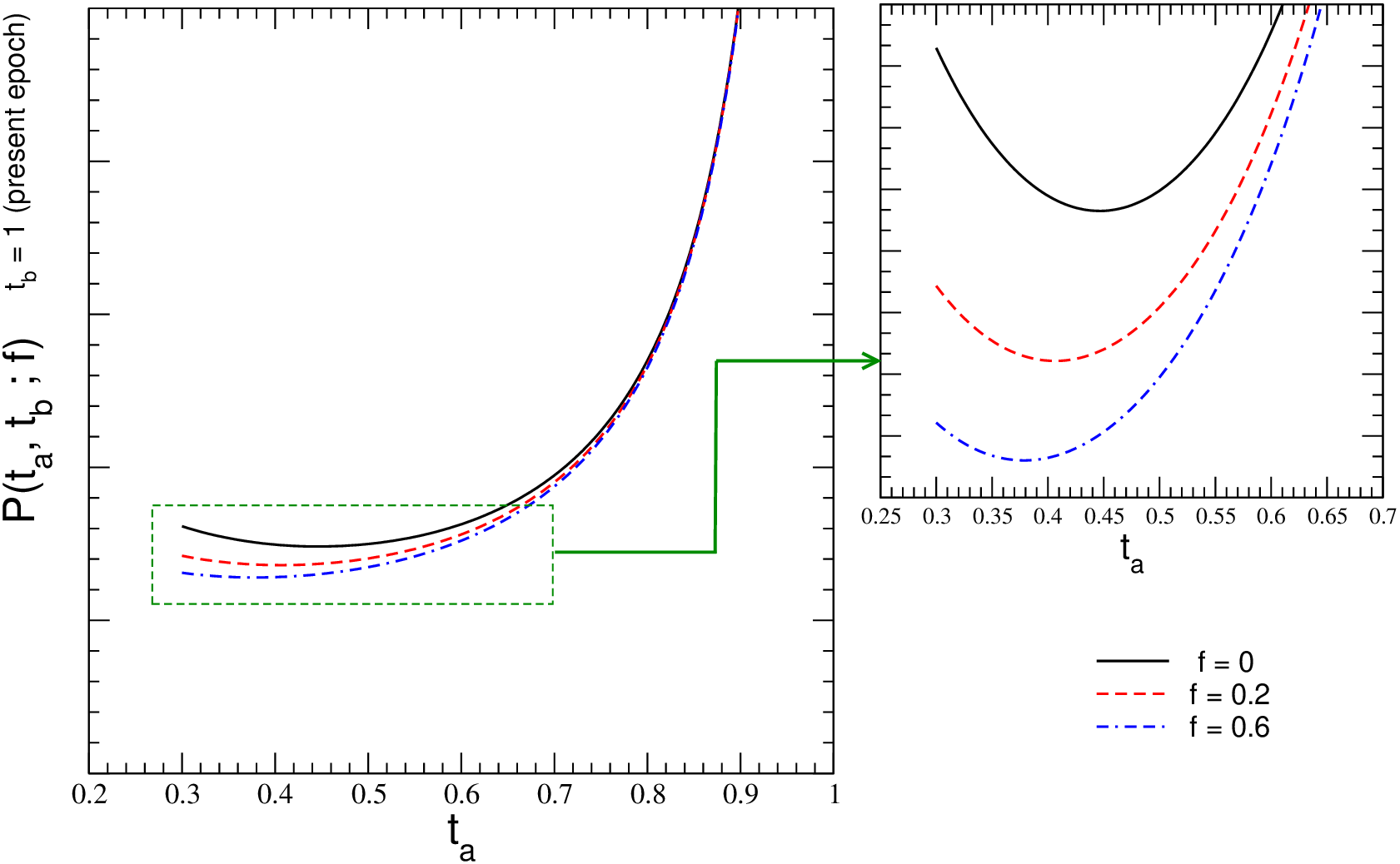}
\caption[]
{Plot of $P(t_a,t_b,f)$ vs $t_a$ for $t_b=1$ (present epoch) for
the homogeneous case and two non-zero benchmark 
values of inhomogeneity parameter
$f$}
\label{fig:4}
\end{figure}
%
\begin{figure}[t]
\centering
\includegraphics[width=\textwidth, height=2.1in]{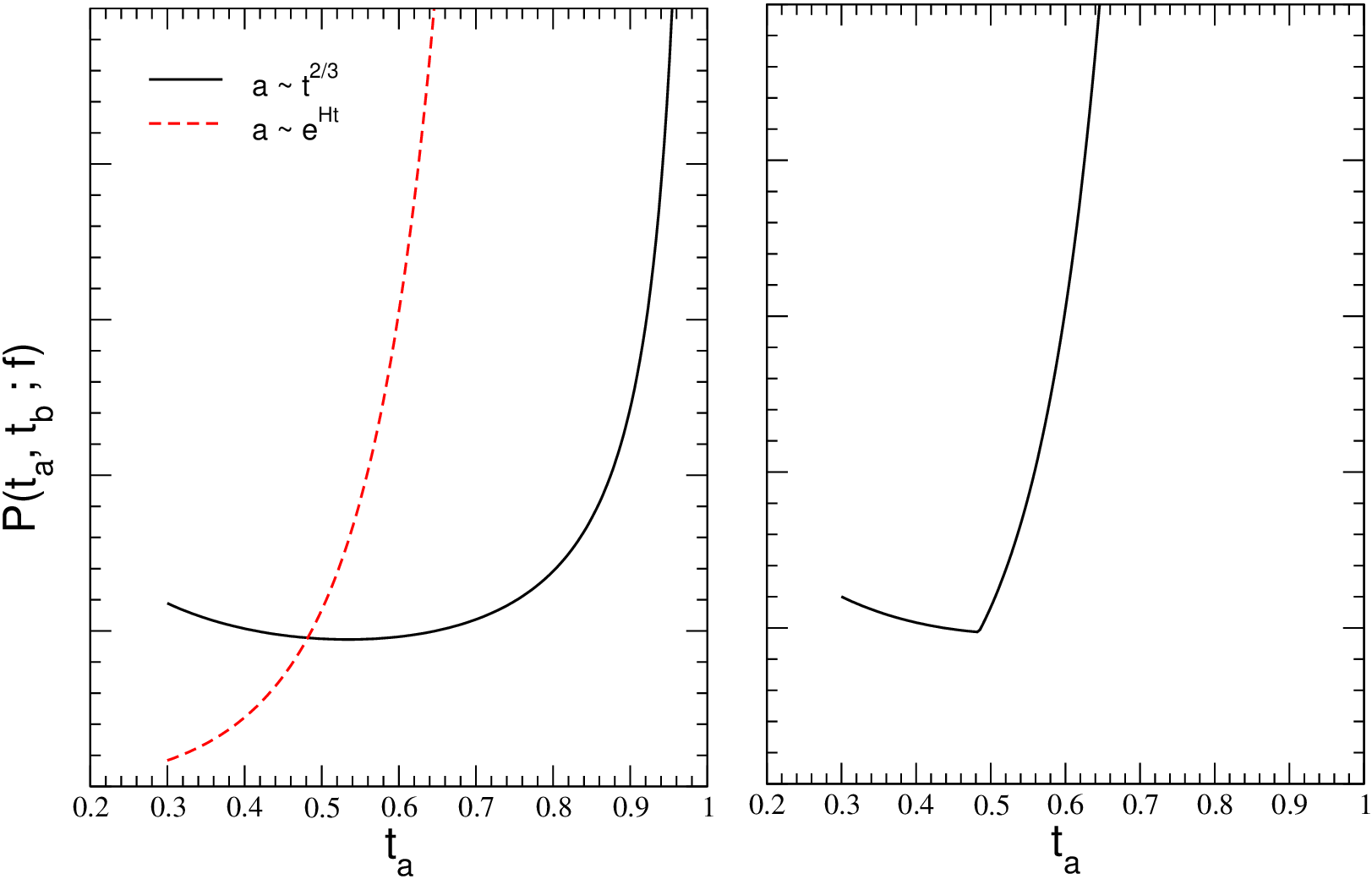}
\caption[]
{Left panel: Plot of $P(t_a,t_b)$ vs $t_a$ for $t_b=1$ (present epoch) for
the homogeneous case with $a(t) \sim t^{2/3}$ for all all epochs (solid line)
and with $a(t) \sim e^{Ht}$ for all epochs (dotted line).
Right panel: Plot of $P(t_a,t_b)$ vs $t_a$ for $t_b=1$ (present epoch) for
the homogeneous case with $a(t) \sim t^{2/3}$ for $t<0.5$ and
with $a(t) \sim e^{Ht}$ for $t>0.5$.}
\label{fig:5}
\end{figure}
%
\begin{figure}[!ht]
\vspace{5mm}
\centering
\includegraphics[width=\textwidth, height=3.0in]{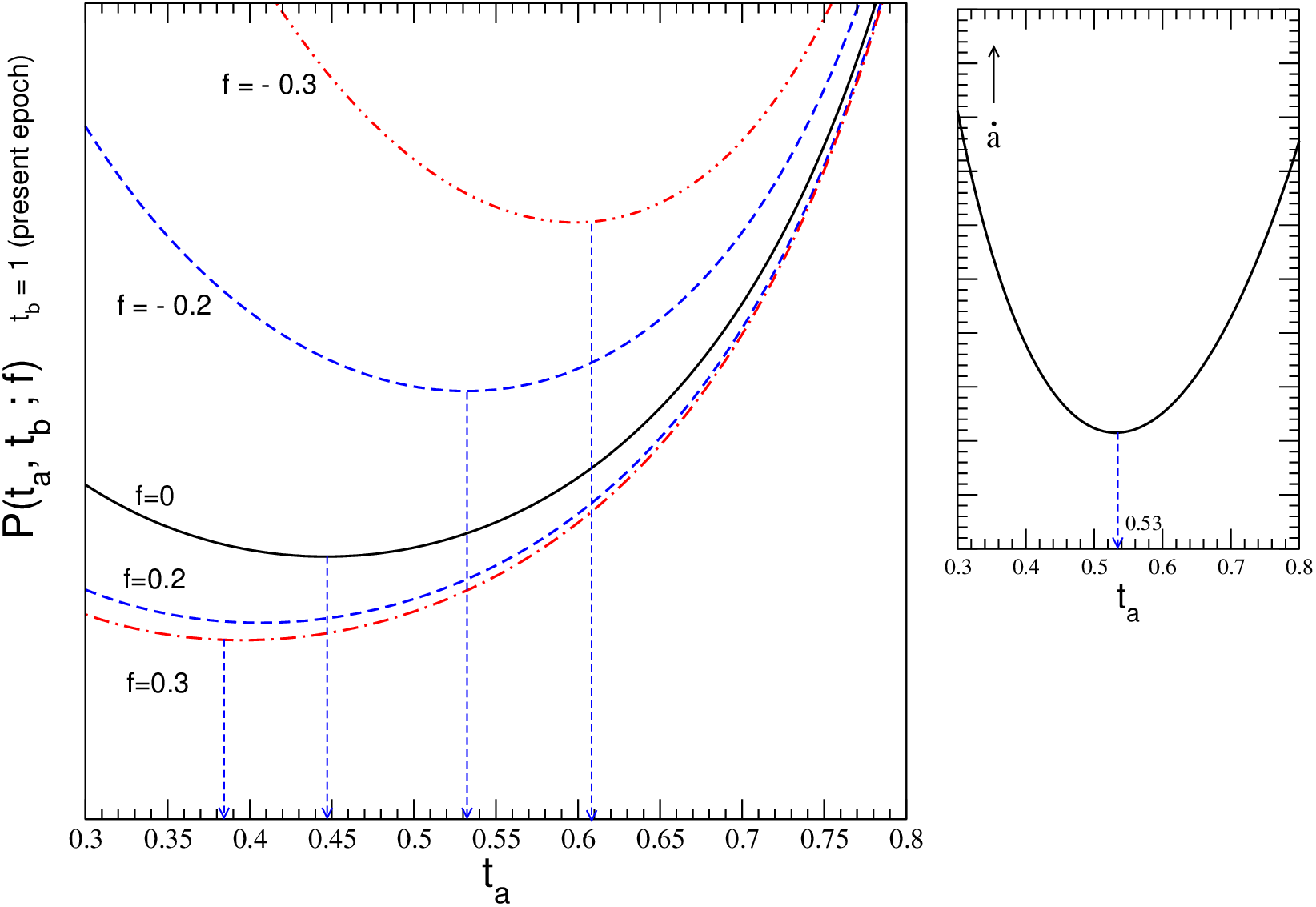}
\caption[]
{Left panel: Plot of $P(t_a,t_b)$ vs $t_a$ for $t_b=1$ (present epoch) for
the different values of $f$, Right panel : Plot of $\dot{a}(t)$ vs $t$ 
obtained from analysis of observational data.}
\label{fig:6}
\end{figure}
We now  investigate various aspects of the transition
probability (Eq.\ (\ref{eq:expand}))  and implications of its  
dependences  on the two parameters of the theory  - 
$\gamma$ (appearing in the solution of Ermakov-Pinney equation)
and $f$ (phenomenological parameter as a measure inhomogeneity), 
as well as the epochs ($t_a,t_b$) between which the transition is considered.
To calculate the transition probability, we use the fact microwave 
background temperature at time 
$t$, $T(t) \propto \frac{1}{a(t)}$ and
we make use of time dependence of $a(t)$ as obtained
from the combined analysis of SNe Ia and observational Hubble data
described in Sec.\ \ref{sec:3}.

To find the $\gamma$ sensitivity of the theory we investigate the
$\gamma$-dependence of the the quantity ${\cal P}(\gamma)$ defined as
\begin{eqnarray}
{\cal P} (\gamma)
& \equiv &
\frac{P_\gamma(t_b,t_a) - P_{\gamma = \epsilon}(t_b,t_a)}
{P_{\gamma = \epsilon}(t_b,t_a)}
\label{eq:calp}
\end{eqnarray}
where we take $\epsilon = 10^{-6}$. For a given $t_a$,$t_b$ the
quantity ${\cal P}(\gamma)$ is a measure of fractional change
in the transition probability  due to variation of the parameter
$\gamma$ from the arbitrarily chosen small value $10^{-6}$.
In figure (\ref{fig:3}) we plot ${\cal P}(\gamma)$ vs $\gamma$,
for fixed values of $t_a$, $t_b$ and taking $f=0$. For all the plots
we have kept fixed $t_b=1$ (present epoch) and have shown the
plots for different chosen values of $t_a$. 
We see from the figure that for small $\gamma$,   \textit{ viz.} $\gamma  < 0.01$,
${\cal P}$ remains effectively same for all epochs implying that
basic results of the model are insensitive to the values of $\gamma$.

The dependence of the transition probability on the epochs 
$t_a$ and $t_b$, between which the transition has been considered is shown in 
Fig.\ \ref{fig:4} where we plot $P_\gamma(t_a,t_b,f)$
against $t_a$ for three chosen values of the phenomenological
parameter $f$ \textit{viz.} 0, 0.2 and 0.6. 
 We have taken $\gamma = 10^{-6}$ for the plots,
but the plots will remain same for all values of $\gamma <0.01$ as 
already seen from results presented in Fig.\ \ref{fig:3}. 

The plots of Fig.\ \ref{fig:4} show that
the probability of transition from an epoch $t$
to the present epoch $t=1$ falls slowly with $t$
in the decelerated phase of expansion of the universe.    
It attains a minimum at a value of $t$ near $t \approx 0.5$  
which is almost the same epoch where the
expansion of universe enters the accelerated phase
from decelerated phase. The probability starts increasing with
$t$ in the accelerated phase of expansion and rises sharply
with $t$ as it approaches more towards the present
epoch. So the nature of $t$-dependence of the 
probability $P(t,t=1)$ has a profile similar to
that of $\dot{a}$, as evident from the middle panel of
 Fig.\ \ref{fig:2}.

The nature of the Fig.\ \ref{fig:4} is further corroborated by an interesting plot.
In the left panel of Fig.\ \ref{fig:5} the solid line is the plot of
$P(t_a,t_b=1)$ vs $t_a$ with the $t-$dependence 
of $a(t)$ taken as $a(t) \sim t^{2/3}$ for all epochs. This corresponds
to a decelerated expansion of the universe dominated by matter. The dotted line is the 
plot of $P(t_a,t_b=1)$ vs $t_a$ with the $t-$dependence 
of $a(t)$ as $a(t) \sim e^{Ht}$ for all epochs. This corresponds
to a dark energy driven accelerated expansion of universe.
The $t_a$-dependence of $P(t_a,t_b=1)$ obtained in
the theory  is given by Eq.\ (\ref{eq:expand}). The $t_a$ dependence enters
in the probability through  a multiplicative term 
$(\ln T_a)^2 \sim (\ln a(t_a))^2 \equiv \eta_1$ (say),
and through the term in  the square bracket $\eta_2$, say.
For low values of $\gamma$ used in our computation,
the term $\eta_2$ increases with $t_a$ as $t_a$ approaches
the present epoch ($t_b=1$),
while the term $\eta_1$ goes as $(\ln t_a)^2$ for $a \sim t^{2/3}$ (matter dominated
universe) and as $t_a^2$ for $a \sim e^{Ht}$ (dark energy dominated universe).
Since the time parameter $t$ we use is normalised to 1 at present epoch,
$t_a$ is fractional, and $(\ln t_a)^2$  decreases as $t_a$ approaches
$t_b=1$, while  $t_a^2$ always increases with $t_a$ in the domain
under consideration. The fact that the probability is a product of $\eta_1$ and $\eta_2$,
there is  a resultant behaviour which determines the
turning point (minimum) at $t_a \sim 0.5$ 
after which the contribution from $\eta_2$
dominates and we are in the accelerated phase.

In the right panel of Fig.\ \ref{fig:5} we  
plot $P(t_a,t_b=1)$ vs $t_a$ with $a(t) \sim t^{2/3}$  up till $t_a \sim 0.5$.
That is, the behaviour of $a(t)$ is taken to be that of a matter dominated
universe. After   $t_a \sim 0.5$ we take the behaviour of $a(t) \sim e^{Ht}$
and compute $P(t_a,t_b=1)$. Here we are taking a dark energy dominated
scenario. Note that the graph mimics  Fig.\ \ref{fig:4} to a great extent except 
for a discontinuous portion around $t_a \sim 0.5$. This is expected
as the change over from a decelerating to an accelerated phase is
bound to be associated with a discontinuity which cannot be
analytically obtained from a phenomenological model.
Mathematically also this is expected because there cannot be a smooth transition 
from $t^{2/3}$ behaviour to that of $e^{Ht}$.
However the overall behaviour (\textit{i.e.} a transition from decelerating
to an accelerating phase) is reflected in the graphical transcription.

From Fig.\ \ref{fig:4} we also get a qualitative 
indication how the presence of inhomogeneity affects the 
evolution of transition probability between different temperatures.
This is shown more comprehensively in Fig.\ \ref{fig:6} where we have 
again plotted $P(t_a,t_b=1)$ vs $t_a$ (left panel) for different 
values of $f$, both positive and negative. For comparison
we have also shown the plot of $\dot{a}(t)$ vs $t$ obtained 
from analysis of observational data in the right panel.
Fig.\ \ref{fig:6} shows that the probability 
is sensitive to the presence of inhomogeneity. This sensitivity 
increases as one goes further into the past and
larger absolute value of the inhomogeneity parameter, $|f|$,
leads to larger departure from probability
values corresponding to the homogeneous scenario.
Also the value of the inhomogeneity parameter determines the epoch of 
switch over from a decelerated to an accelerated phase of expansion
of the universe. A positive value leads to 
switch over at earlier epochs, while a negative value leads to 
switch over at later epochs. We obtain better agreement 
with the observed value of crossover point $t \approx 0.53$ 
for a negative value of inhomogeneity parameter $f \approx -0.2$.

\section{Conclusion}
In this work a phenomenological model has been developed to study the evolution 
of the universe in the context of the CMBR. A key ingredient of the model is the 
presence of dark energy through a scalar field whose kinetic energy dominates i.e. 
a $k-$essence scalar field. We first develop the observational evidence through 
a rigorous graphical transcription of SNe Ia data. This is depicted in 
Fig.\ \ref{fig:2}. Subsequently a Lagrangian model of dark energy 
(obtained from very general considerations, Sec.\ \ref{sec:2}) is used to explain 
the evidence depicted in Fig.\ \ref{fig:2}.  

The approach taken is as follows. The Lagrangian (Eq.\ (\ref{eq:lagosc})) is that of a time 
dependent oscillator and the dynamical variables is $q=\ln a(t)$. 
We compute the  the quantum fluctuations $\langle q_{a}, t_{a}|q_{b},t_{b}\rangle$ 
which is tantamount to computing the correlations between the logarithm of the 
temperatures at two epochs $t_{a}$ and $t_{b}$  where we have used the association  
between the scale factor $a(t)$ and cosmic temperature at a particular epoch, 
{\textit viz.}, $T_{a}\sim {\frac{1}{a(t_{a})}}$.  

What is remarkable is that the probability of transition between 
the logarithm of the temperatures   
$\ln T_{a}$ at $t=t_{a}$  and  $\ln T_{b}$ at $t=t_{b}\mbox{(present epoch)}$  
follows a similar profile  as that of $\dot a(t)$. 
This is shown in Fig.\ \ref{fig:4}. Another point of note is that the crossover 
from a decelerating phase to an accelerated phase occurs at precisely 
at the same epoch, {\textit viz.}, i.e. $t_{a}=0.5$.

This shows that our model captures the essential physics because the probability 
seems to be proportional to $\dot a(t)$ and as  $a(t + dt)\sim a(t) + \dot a(t) dt$, 
therefore the probability of transition should be proportional to $\dot a$. 
Figs. \ref{fig:2} and \ref{fig:4}
seem to confirm this fact. 

Our phenomenological model successfully explains the  observed variation of  $\dot a$. 
This conclusion follows from the fact that the variation of $\dot a$ 
matches with the probability profile obtained 
theoretically from the model after plugging in observed values of 
$\dot a$ at corresponding epochs. Moreover,
the observed value of the epoch when the  universe went from a 
decelerating phase to an accelerated phase, it is nearly the same 
as that obtained from the theoretically obtained profile.

This model also throws light on how inhomogeneity may affect the 
CMBR evolution. There is a qualitative indication that the probability 
is sensitive to the presence of inhomogeneity and this sensitivity 
increases as one goes further into the past. Also
the value of the inhomogeneity parameter determines the epoch of 
switch over to an accelerated phase. A positive value leads to 
switch over at earlier epochs, while a negative value leads to 
switch over at later epochs. Better agreement with the observed 
value of crossover point is obtained for a small negative value of 
the inhomogeneity parameter. This is seen from Fig.\ \ref{fig:6}.


\end{document}